\documentclass[11pt,twoside]{article}
\usepackage{amsmath}
\usepackage{amssymb}
\usepackage{cite}
\usepackage{graphicx}
\usepackage[english]{babel}

\begin{document}
\newcommand{\pst}{\hspace*{1.5em}}

\newcommand{\rigmark}{\em Journal of Russian Laser Research}
\newcommand{\lemark}{\em Volume 30, Number 5, 2009}

\newcommand{\be}{\begin{equation}}
\newcommand{\ee}{\end{equation}}
\newcommand{\bm}{\boldmath}
\newcommand{\ds}{\displaystyle}
\newcommand{\bea}{\begin{eqnarray}}
\newcommand{\eea}{\end{eqnarray}}
\newcommand{\ba}{\begin{array}}
\newcommand{\ea}{\end{array}}
\newcommand{\arcsinh}{\mathop{\rm arcsinh}\nolimits}
\newcommand{\arctanh}{\mathop{\rm arctanh}\nolimits}
\newcommand{\bc}{\begin{center}}
\newcommand{\ec}{\end{center}}

\newcommand{\aver}[1]{\left< #1 \right>}

\thispagestyle{plain}

\label{sh}


\begin{center} {\Large \bf
\begin{tabular}{c}
DYNAMICS OF A NONLINEAR QUANTUM OSCILLATOR 
\\[-1mm]
UNDER NON-MARKOVIAN PUMPING
\end{tabular}
 } \end{center}

\bigskip

\bigskip

\begin{center} {\bf
A.~D.~Alliluev and D.~V.~Makarov$^*$
}\end{center}

\medskip

\begin{center}
{\it
$^1$Laboratory of Nonlinear Dynamical Systems, \\ V.~I.~Il'ichev Pacific Oceanological Institute\\
Far East Branch of the Russian Academy of Sciences\\
Vladivostok, Russia 690041}

\smallskip

$^*$Corresponding author e-mail:~~~makarov~@~poi.dvo.ru\\
\end{center}

\begin{abstract}\noindent

We consider dynamics of a quantum nonlinear oscillator subjected to non-Markovian pumping.
Models of this kind can describe formation of exciton-polariton Bose-Einstein condensates in course of stimulated scattering
and relaxation of reservoir excitons.
Using the Markovian embedding techniques, we obtain stochastic differential equations of motion with an additional degree of freedom corresponding to dynamical memory.
It is shown that the oscillator asymptotically tends to the 
intrinsically non-Markovian stable fixed point corresponding to constant product
of oscillator amplitude and modulo of the memory variable. The state corresponding to this point exhibits unlimited growth of population,
with the growth rate that decreases with time.
Our results show that 
the Markovian behavior could be observed only within some limited early stage of oscillator evolution 
provided that decay of dynamical memory is sufficiently fast.
Transition from the Markovian regime to non-Markovian one with increasing time is linked to phase shift of the pumping term.
Coherence properties of the oscillator are studied. It was found that interaction between particles delimits growth of condensate population,
thereby impeding formation of Bose-Einstein condensate.
\end{abstract}

\medskip


\section{Introduction}
\pst

When we 
study quantum systems which interact with environment, we commonly simplify the problem by 
eliminating  multiple environmental degrees of freedom.
In this way, a quantum system ``of interest'' is referred to as an open quantum system, whereas environment
is treated as some external reservoir.
Interaction of quantum systems with thermal reservoirs is one of the most important problems
in contemporary physics. It is of vital importance in the context of
physics of cold and ultracold quantum gases \cite{Proukakis,Prants-PRA,Prants-JETPL,Prants-JRLR,Bychek}, 
quantum computation \cite{QC}, quantum fluids with a photonic component \cite{Deng,Carusotto},
atomtronics \cite{Atomtronics} and many other fields \cite{BP} where 
finite-temperature effects play a significant role.

Assumption of reservoir independence on dynamics of the open quantum system naturally leads us to the Markov approximation. 
This assumption is valid if the reservoir is characterized by relatively  broad energy spectrum (``hot'' reservoir) 
and/or fast decay of dynamical correlations (``chaotic'' reservoir). The Markov approximation yields time-local equations governing quantum evolution.
However, recent progress in experimental technologies has led to onset of quantum systems where 
the role of a reservoir is played by subsystems with narrow energy spectra and high degree of coherence.  
Examples of such systems are optically trapped atoms with the Raman coupling to ballistic states \cite{Hope,deVega08,Navarette-Benloch,Krinner},
quantum dots, qubits or nanomechanical oscillators coupled to Bose-Einstein condensate (BEC) \cite{Recati,Sokolovski,Hunger,Apollaro,Lampo}, motion 
of matter-wave solitons experiencing friction caused by excitations of a superfluid \cite{Efimkin-PRL}, 
quantum systems coupled to spin baths \cite{Prokofev}.
If a reservoir cannot be fully separated from a considered quantum state,
dynamics becomes essentially non-Markovian.
Non-Markovian open quantum systems are 
basically governed by evolution equations which are non-local in time and contain integrals with memory kernels 
\cite{Diosi97,Strunz99,Yu,Adamian,Chakraborty,Schiro}.
Inclusion of memory effects can remarkably enrich physical properties 
of such systems (see \cite{Ribeiro_Vieira,Breuer-RMP,deVega} for particular illustrations).

Most studies of non-Markovian open quantum systems consider the case of non-Markovian dissipation. The case of non-Markovian pumping
is relatively poorly studied. Pumping of this kind can occur under stimulated relaxation of particles belonging to a reservoir.
A particular example is exciton-polariton BEC interacting with excitonic reservoir \cite{EL18,MEL}. Incoherent pumping of the reservoir density can enhance condensate formation whereby improving coherence properties of the outcoming light, this phenomenon
can be used for development of polaritonic lasers \cite{Kim}.
Also, one should expect emergence of non-Markovian features under evaporative cooling of atoms loaded into a dipole optical trap \cite{Ronen}.

In the present paper we consider dynamics of a quantum oscillator describing
a single state experiencing pumping from a Lorentzian bath in the non-Markovian regime. 
In the mean-field approximation for interaction of particles,
state dynamics is governed by a kind of the nonlinear Schr\"odinger equation.
Although this work is motivated by physics of exciton-polariton BEC, a paradigmatic form of the model considered anticipates that
it can be used in a variety of physical problems.
We concentrate attention on nonlinear phenomena in dynamics of the oscillator.

The paper is organized as follows. 
In the next section we describe the model under study.
Section \ref{sec:Stationary} is devoted to stationary solutions for the oscillator.
Coherence of the oscillator is studied in Sec.~\ref{sec:Coherence} in the context of Bose-Einstein condensate formation.
The main results of the paper are discussed in Summary.

\section{The model}
\label{Model}

Let's consider an ensemble of cold bosonic particles, where small fraction belongs to some single state, while the remaining part is characterized by narrow but continuous energy spectrum and can be considered as precondensate.
If number of particles is sufficiently large, we can
invoke the semiclassical approximation, when
amplitude of the state can be represented by a c-number \cite{Kolovsky-PRA81,JRLR,CNSNS,Richaud,Iubini_Lepri_Livi_Politi,Iubini_Politi,EPJB20}.
Under certain conditions (which shall be discussed in Sec.~\ref{sec:Coherence}) particles belonging to the single
state can form condensate. 
Considering the precondensate as reservoir and 
utilizing the mean-field approximation,
we can describe evolution of the single state amplitude by means of the following stochastic nonlinear Schr\"odinger equation:
\begin{equation}
 i\hbar\frac{da}{dt} = E_0a + \alpha_{\text{int}}|a|^2a + \eta(t) + \hbar\int dt' \Sigma(t,t') a(t'),
 \label{sys0}
\end{equation}
where $E_0$ is state energy, $\alpha_{\text{int}}$ is a nonlinearity rate quantifying 
strength of particle interaction, 
 $\Sigma(t,t')$ is a retarded self-energy utilized to describe coupling to the reservoir,
and $\eta$ describes fluctuations induced by stochastic transitions between the state and the reservoir.
In this model we neglect interaction of precondensate particles, assuming that their phase space density is low.

Model (\ref{sys0}) can be referred to as a quantum nonlinear oscillator under non-Markovian pumping.
The self-energy can be expressed as \cite{EL18,MEL}
\begin{equation}
 \Sigma(t,t') = 2iV^2 \rho^2 K(t,t'),
\end{equation}
where $V$ is rate of coupling between the oscillator and the reservoir, 
and $\rho$ is reservoir population.
Form of the memory kernel $K(t,t')$ depends on energy spectrum of the reservoir \cite{Fraga}. 
In the present work we consider the Lorentzian spectrum 
\begin{equation}
 S(E) = \frac{\gamma}{[(E-E_{\text{c}})/\hbar]^2 + \gamma^2},
 \label{Lorentz}
\end{equation}
that yields 
\begin{equation}
 K(t,t') = \frac{\gamma}{2} e^{-\gamma(t-t')},
\end{equation}
for $E_{\text{c}} = E_0$. 
One can see that the parameter $\gamma$ simultaneously determines spectral width of the reservoir and 
rate of memory decay.
However, if 
\begin{equation}
 \Omega \equiv \frac{E_{\text{c}} - E_0}{\hbar} \ne 0,
 \label{detuned}
\end{equation}
then the memory kernel is supplemented by an oscillating factor and becomes \cite{EL18}
\begin{equation}
 K(t,t') = \frac{\gamma}{2} e^{-(\gamma + i\Omega)(t-t')}.
 \label{Ktt}
\end{equation}
Population of the reservoir obeys equation

\begin{equation}\label{rhoPump}
    \frac{d\rho}{dt} = F(t) - 2\gamma_{\text{R}}\rho(t) - \frac{2}{\hbar}\text{Re}\left[a^{*}(t)\eta(t)\right]- 2\text{Re}\left[a^{*}V^2\rho^2\int\limits_{t'=0}^{t}K(t,t')a(t')\,dt' \right],
\end{equation}
where function $F(t)$ describes incoherent pumping of the reservoir, 
$\gamma_{\text{R}}$ is decay rate for reservoir particles. 
In experiments with exciton-polaritons, incoherent pumping is produced by external laser radiation.
We consider the simplest model of the incoherent pumping, when $F=\text{const}$. 
In the absence of particle exchange between the reservoir and the oscillator, Eq. (\ref{rhoPump}) can be easily integrated that gives

\begin{equation}
    \rho(t) = \rho(0)e^{-2\gamma_{\text{R}}t}+\frac{F}{2\gamma_{\text{R}}}(1-e^{-2\gamma_{\text{R}}t}).
\end{equation}
In the limit $t\rightarrow \infty$ we have

\begin{equation}\label{incoherPump}
    \rho(t)\rightarrow\rho_{0}=\frac{F}{2\gamma_{\text{R}}}.
\end{equation}
Substituting (\ref{incoherPump}) into (\ref{rhoPump}), we get

\begin{equation}
 \frac{d\rho}{dt} = 2\gamma_{\text{R}}(\rho_0 - \rho) - \frac{2}{\hbar}\text{Re}\left[a^{*}(t)\eta(t)\right]- 2\text{Re}\left[a^{*}V^2\rho^2\int\limits_{t'=0}^{t}K(t,t')a(t')\,dt' \right],
 \label{drhdt}
\end{equation}
The constant $\rho_0$ determines the equilibrium reservoir population 
in the absence of coupling to the oscillator.
Inserting (\ref{Ktt}) into (\ref{sys0}) and (\ref{drhdt}), we obtain
\begin{equation}
 i\hbar\frac{da}{dt} = E_0a + \alpha_{\text{int}}|a|^2a + \eta(t) + i\hbar\gamma V^2\rho^2\int\limits_{t'=0}^{t} e^{-(\gamma + i\Omega)(t-t')}a(t')\,dt',
 \label{sys1}
\end{equation}

\begin{equation}
 \frac{d\rho}{dt} = 2\gamma_{\text{R}}(\rho_0 - \rho) - \frac{2}{\hbar}\text{Re}\left[a^{*}(t)\eta(t)\right]- \text{Re}\left[a^{*}V^2\rho^2\gamma\int\limits_{t'=0}^{t}e^{-(\gamma + i\Omega)(t-t')}a(t')\,dt' \right],
 \label{drhdt1}
\end{equation}
It should be mentioned that this model can be considered a particular case
of the Stuart-Landau oscillator \cite{Selivanov,Hohlein,Kemeth,Hongjie}.

Autocorrelation function of fluctuations $\eta(t)$ is determined by the Keldysh component of self-energy.
It is linked to the memory kernel via the fluctuation-dissipation relation
\begin{equation}
 \aver{\eta^*(t)\eta(t')} = 2\hbar^2V^{2}\rho^2(t) K(t,t')=\hbar^2\gamma V^{2}\rho^2(t) e^{-(\gamma + i\Omega)(t-t')}.
 \label{FDT}
\end{equation}
It corresponds to one-dimensional complex-valued Ornstein-Uhlenbeck process \cite{Arato} 
described by the following Langevin equation

\begin{equation}
    \dot \eta(t) = -2(\gamma+i\Omega)\eta(t) + \sqrt{2\gamma}\xi(t).
\end{equation}
Here $\xi(t)$ is Gaussian complex-valued white noise with $\aver{\xi(t)}=0$, $\aver{\xi(t)\xi^{*}(t')}=\delta(t-t')$.
If $\gamma\ll\Omega$, then $\eta(t)$ can be fairly approximated by
the complex-valued harmonic noise introduced in \cite{PLA,EPJB14,PhysScr,Zitterbewegung,QE}. 

Using the Markovian embedding method \cite{deVega,MEL,Xiantao}, we can remove integrals from the right-hand sides of Eqs. (\ref{sys1}) and (\ref{drhdt1}).
It can be obtained by introducing the memory variable defined as \cite{Neiman}
\begin{equation}
 M = a_0e^{-(\gamma + i\Omega)t} + \gamma \int\limits_{t'=0}^{t} e^{-(\gamma + i\Omega)(t-t')}a(t')\,dt',
 \label{Mdef}
\end{equation}
where $a_0=a(t=0)$.
Then Eqs. (\ref{sys1}) and (\ref{drhdt1}) become
\begin{equation}
 i\hbar\frac{da}{dt} = E_0a + \alpha_{\text{int}}|a|^2a + \eta(t) + i\hbar V^{2}\rho^2\left[M-a_0e^{-(\gamma + i\Omega)t}\right],
 \label{sys2}
\end{equation}
and
\begin{equation}
 \frac{d\rho}{dt} = 2\gamma_{\text{R}}(\rho_0 - \rho) - \frac{2}{\hbar}\text{Re}\left[a^{*}(t)\eta(t)\right]- 
 \text{Re}\left(a^{*}V^2\rho^2\left[M-a_0e^{-(\gamma + i\Omega)t}\right] \right).
 \label{drhdt2}
\end{equation}
The memory variable obeys the following equation of motion:
\begin{equation}
 \frac{dM}{dt} = \gamma(a - M) - i\Omega M.
 \label{dMdt}
\end{equation}

\section{Stationary solutions}
\label{sec:Stationary}

\begin{figure}[!ht]
\centering
\includegraphics[width=.32\textwidth]{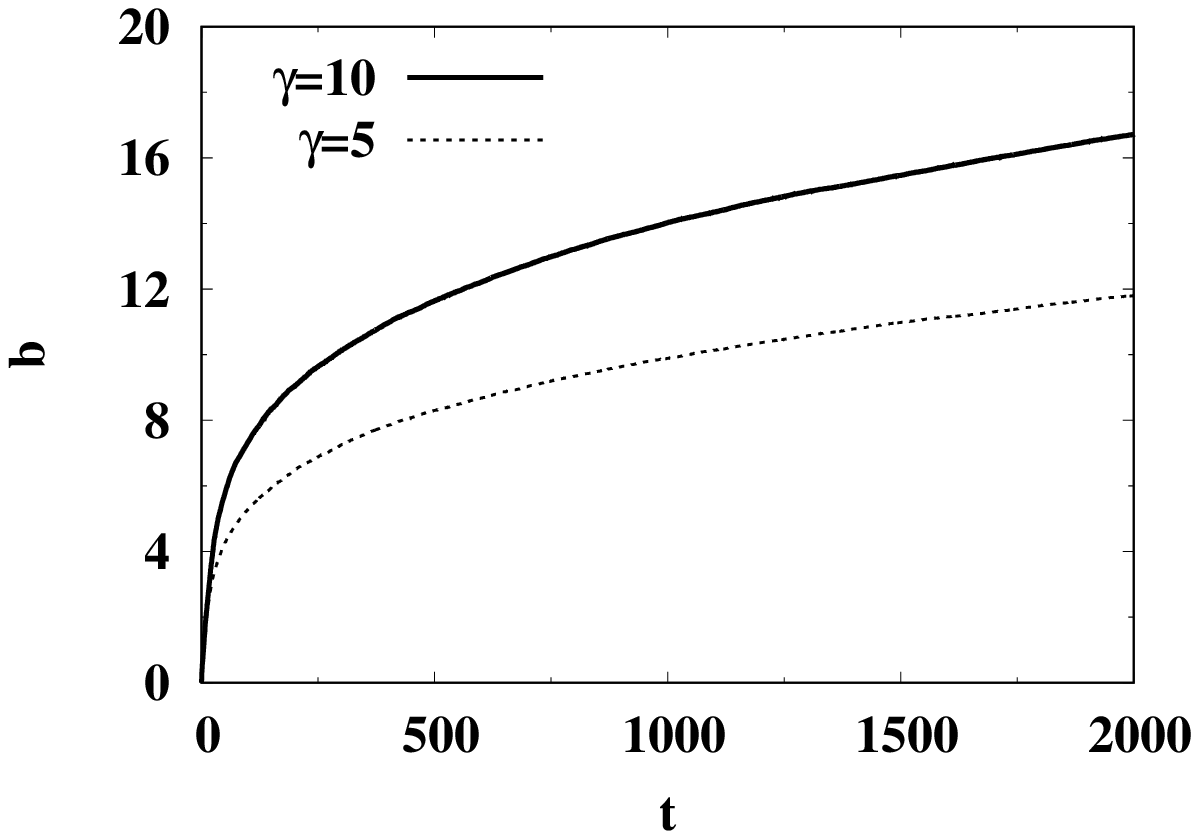}
\includegraphics[width=.32\textwidth]{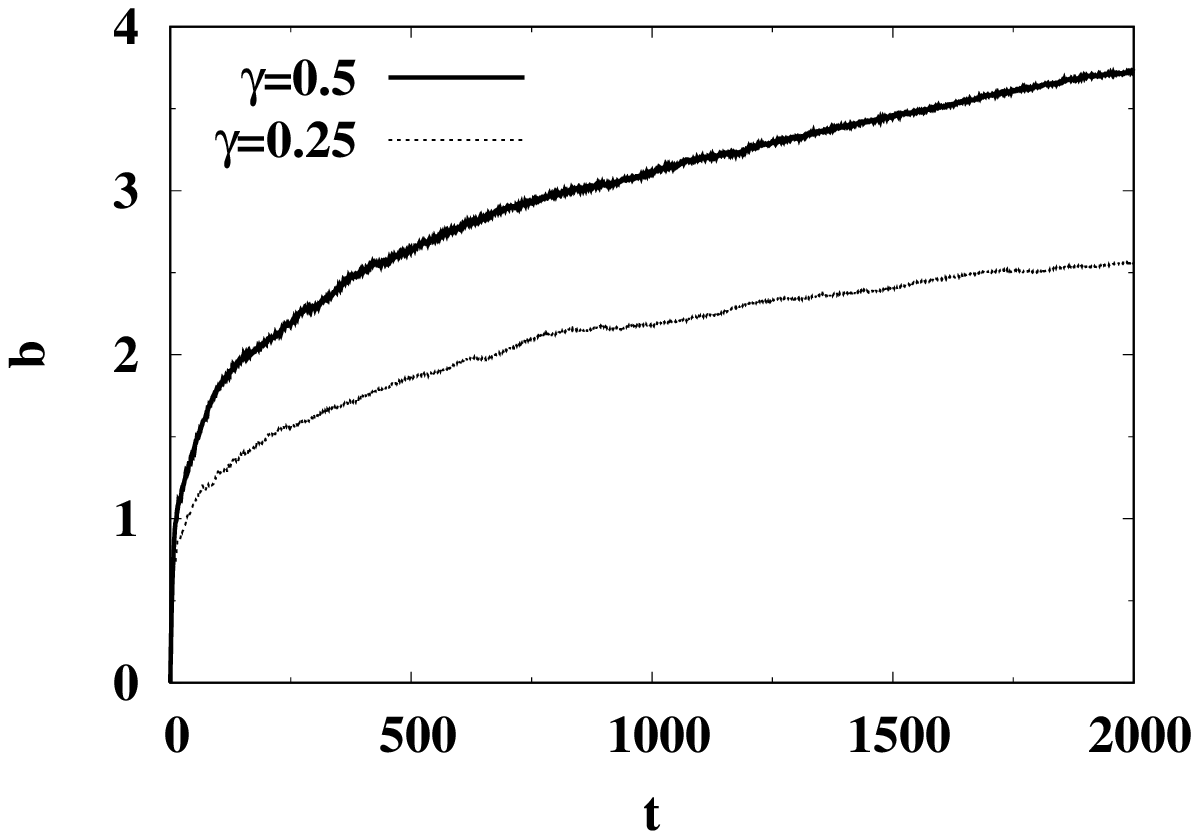}
\includegraphics[width=.32\textwidth]{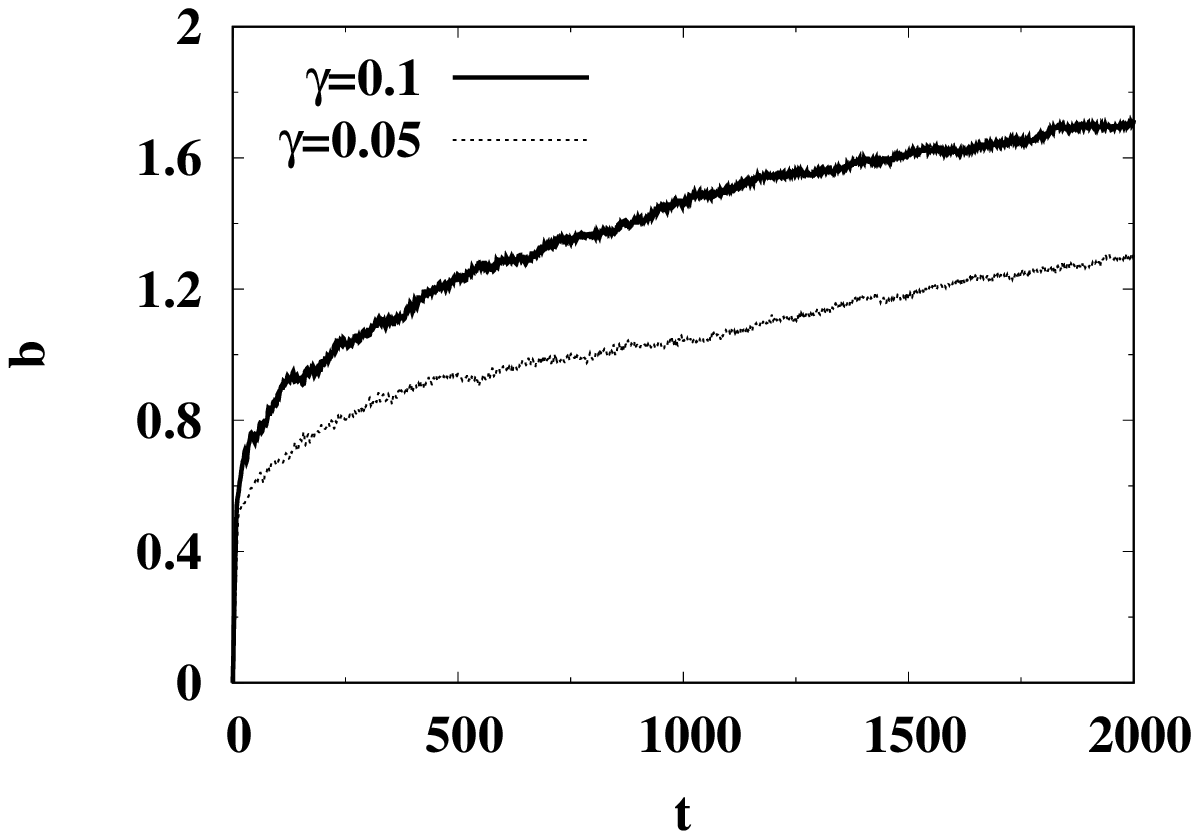}
\caption{Time dependence of oscillator amplitude $b$ averaged over 100 realizations of fluctuations. Parameter values: $\hbar=1, \alpha_{\text{int}}=1, \gamma_R=0.1, \Delta E=0, \rho_0=1000,  V=\sqrt{0.1}/\rho_0 $.
} 
\label{Fig:b_t}
\end{figure} 

Equations (\ref{sys2})--(\ref{dMdt}) form a dissipative dynamical 
system whose long-term dynamics is determined by its stationary manifolds \cite{Gavrilov-UFN,Kolovsky}.
Indeed, there cannot be strictly stationary states in the presence of noise.
However, we consider the case of $\gamma\ll 1$, then noise can be considered as a weak perturbation imposed
onto deterministic dynamics governed by equations of motion without noises. 
Removing noises from the right-hand sides and 
using the substitutions
\begin{equation}
 a(t)=b(t)e^{-i\left(\phi(t)+\frac{E_0}{\hbar}t\right)},\quad
 M = m(t)e^{-i\psi(t)},
\end{equation}
we can rewrite
Eqs. (\ref{sys2})--(\ref{dMdt}) as
\begin{equation}
  \frac{db}{dt}=V^{2}\rho^{2} m\cos{\theta}
\label{dbdt}
\end{equation}
\begin{equation}
     \frac{d\phi}{dt}=\frac{\alpha_{\text{int}}}{\hbar}b^{2}-V^{2}\rho^{2}\frac{m}{b}\sin{\theta}
\end{equation}
\begin{equation}
    \frac{dm}{dt}=\gamma(b\cos{\theta}-m),
    \label{dmdt}
\end{equation}
\begin{equation}
   \frac{d\psi}{dt}=\gamma\frac{b}{m}\sin{\theta}+\frac{\Delta E}{\hbar},
    \label{dpsi}
\end{equation}

\begin{equation}
\frac{d\rho}{dt}=2\gamma(\rho_{0}-\rho)-2bmV^{2}\rho^{2}\cos{\theta},
\label{drho}
\end{equation}
  where $\theta=\phi-\psi$, $\Delta E=\hbar\Omega-E_0$ is
   energy detuning between the reservoir and the oscillator. 
  The exponentially 
decaying term $i\hbar V^2\rho^2 a_0e^{-(\gamma + i\Omega)t}$ in (\ref{sys2}) is also dropped out.

One can see that Eqs. (\ref{dbdt})-(\ref{drho}) do not yield non-trivial stationary states obeying to the condition
\begin{equation}
 \frac{db}{dt} = \frac{dm}{dt} = \frac{d\theta}{dt} = \frac{d\rho}{dt} = 0.
\end{equation}
Their absence anticipates unbounded growth of the oscillator amplitude $b$, that is confirmed by results of numerical simulation presented in Fig.~\ref{Fig:b_t}.
It should be noticed that rate of the growth diminishes with time.
\begin{figure}[!ht]
\centering
\includegraphics[width=.32\textwidth]{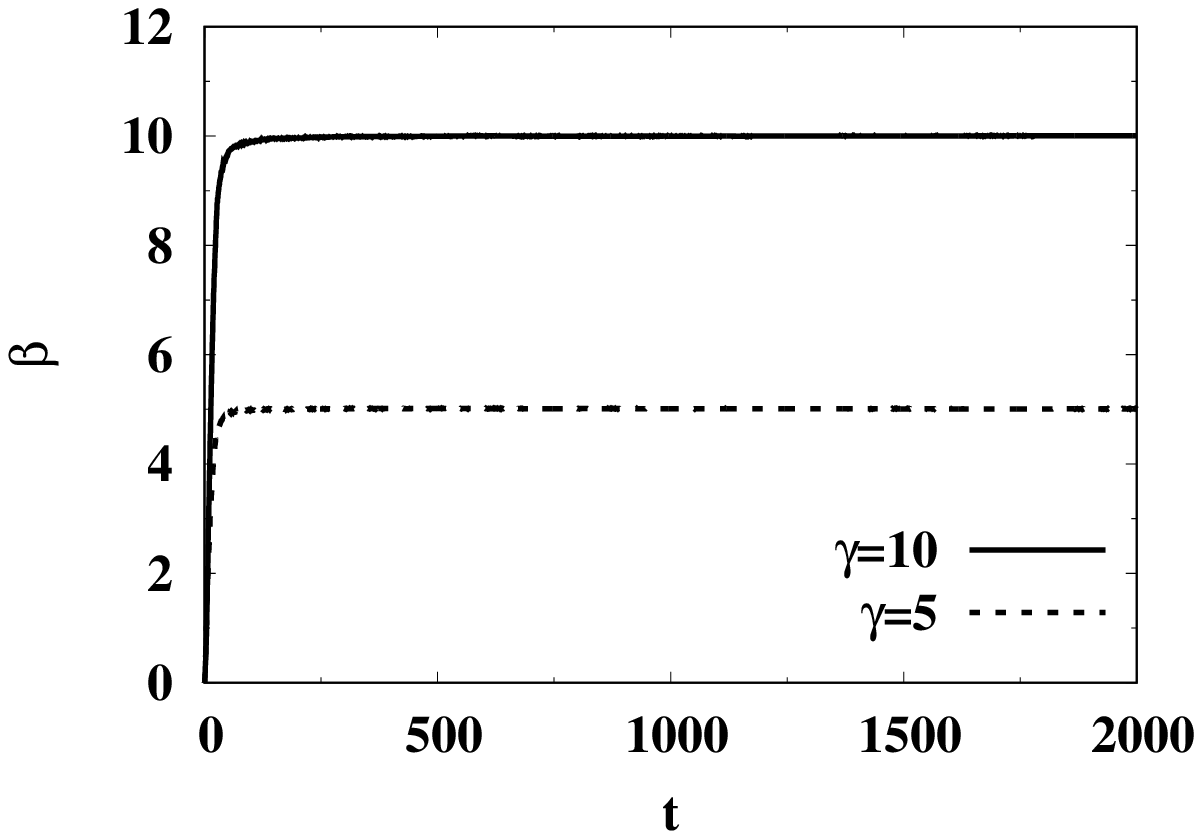}
\includegraphics[width=.32\textwidth]{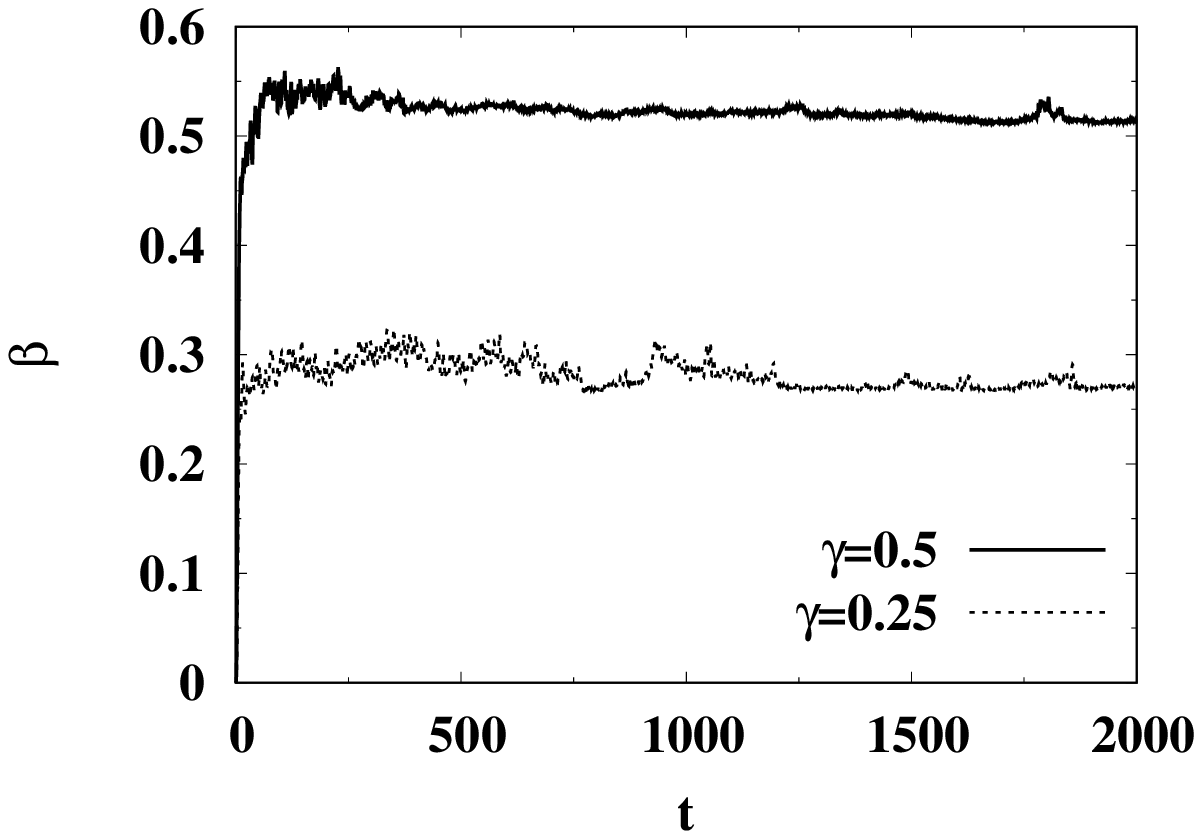}
\includegraphics[width=.32\textwidth]{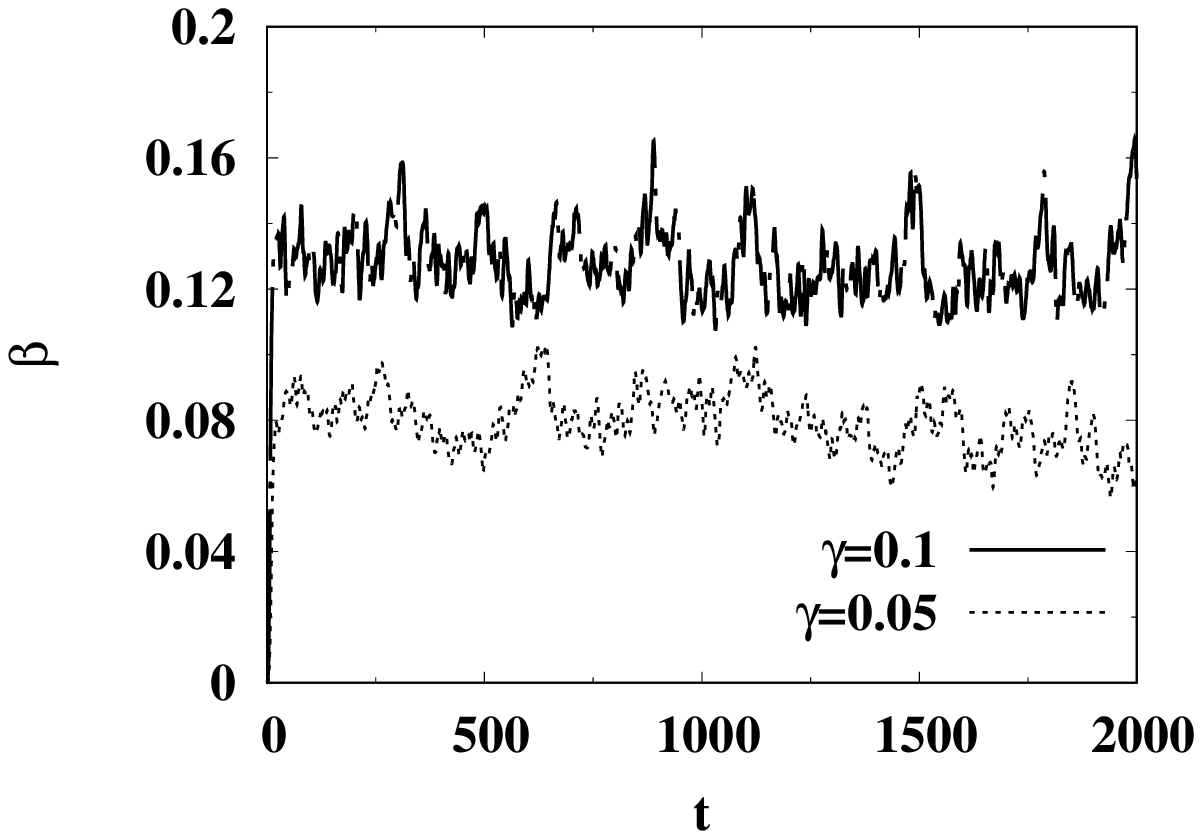}
\caption{Time dependence of combined amplitude variable $\beta$ averaged over 100 realizations of fluctuations.
 Parameter values: $\hbar=1, \alpha_{\text{int}}=1, \gamma_R=0.1, \Delta E=0, \rho_0=1000, V=\sqrt{0.1}/\rho_0 $.
}  
\label{Fig:beta_t}
\end{figure} 

\begin{figure}[h]
\begin{minipage}[h]{0.47\linewidth}
\center{\includegraphics[width=1\linewidth]{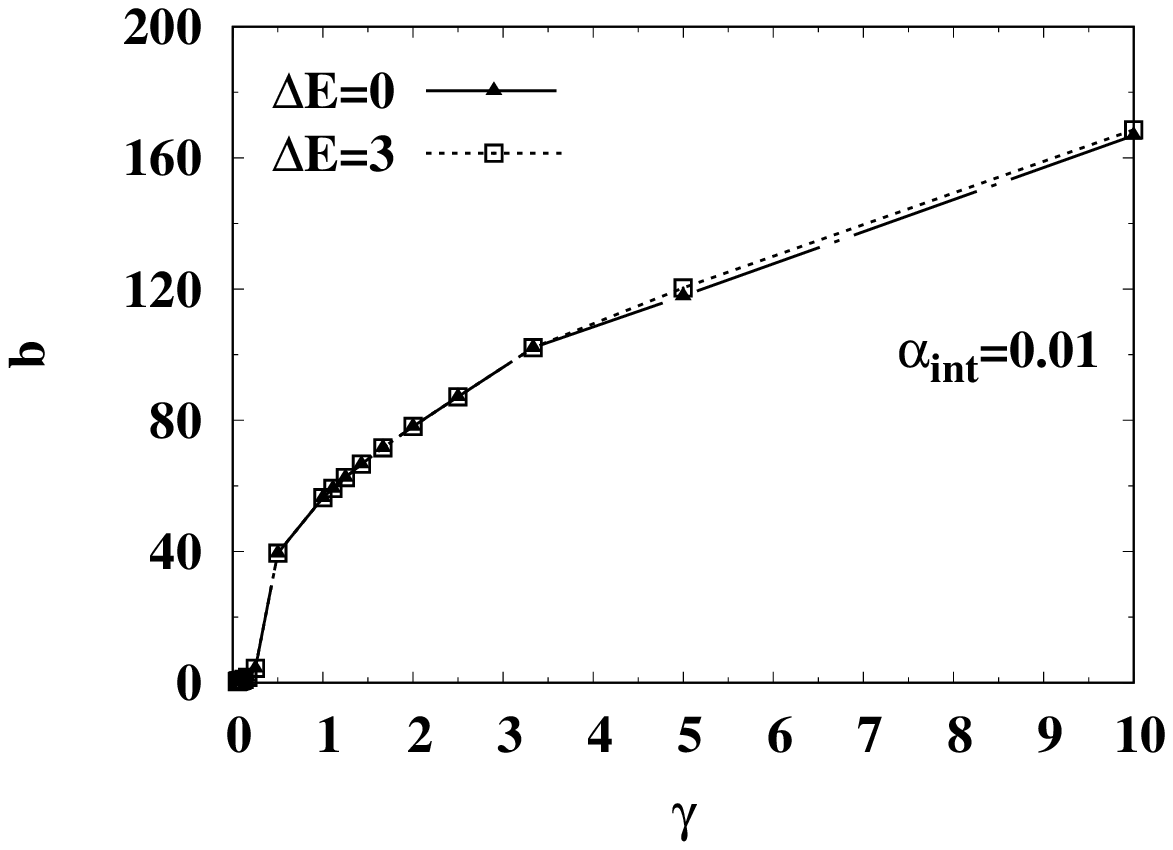}}  \\
\end{minipage}
\hfill
\begin{minipage}[h]{0.47\linewidth}
\center{\includegraphics[width=1\linewidth]{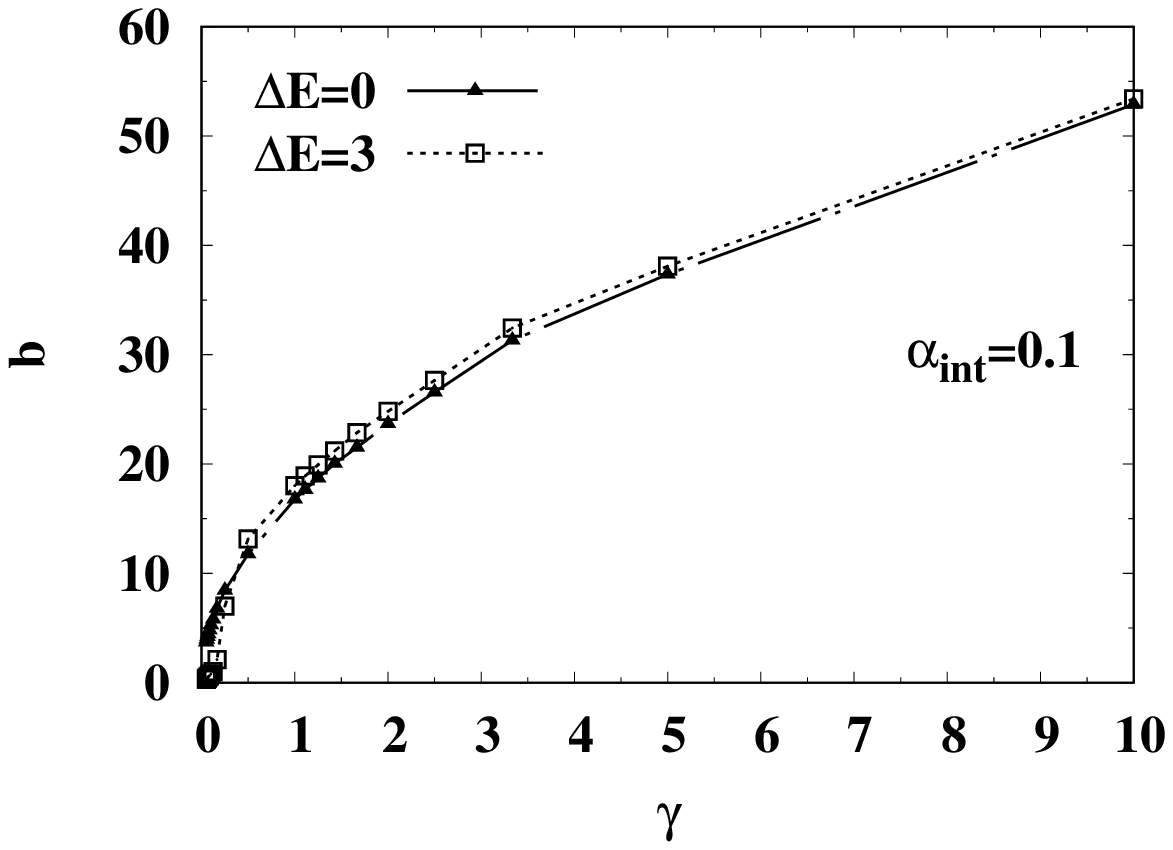}} \\
\end{minipage}
\vfill
\begin{minipage}[h]{0.47\linewidth}
\center{\includegraphics[width=1\linewidth]{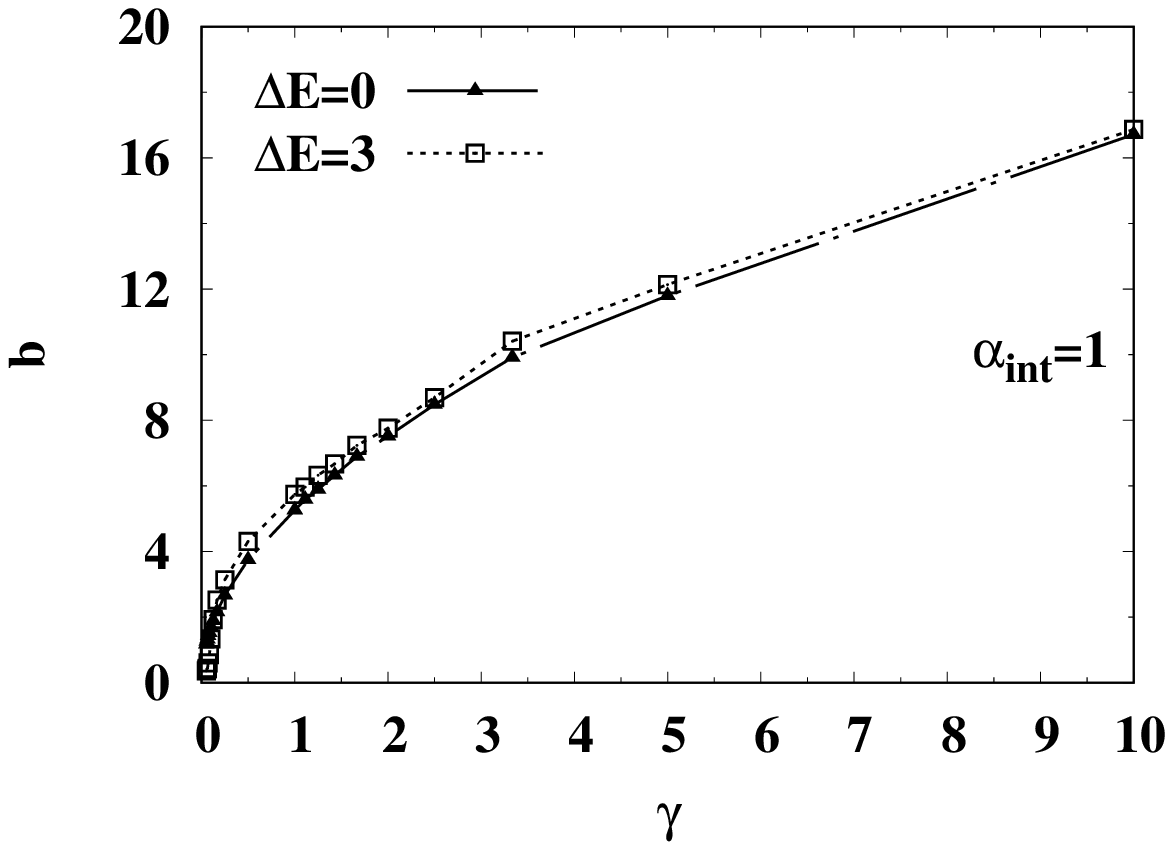}} \\
\end{minipage}
\hfill
\begin{minipage}[h]{0.47\linewidth}
\center{\includegraphics[width=1\linewidth]{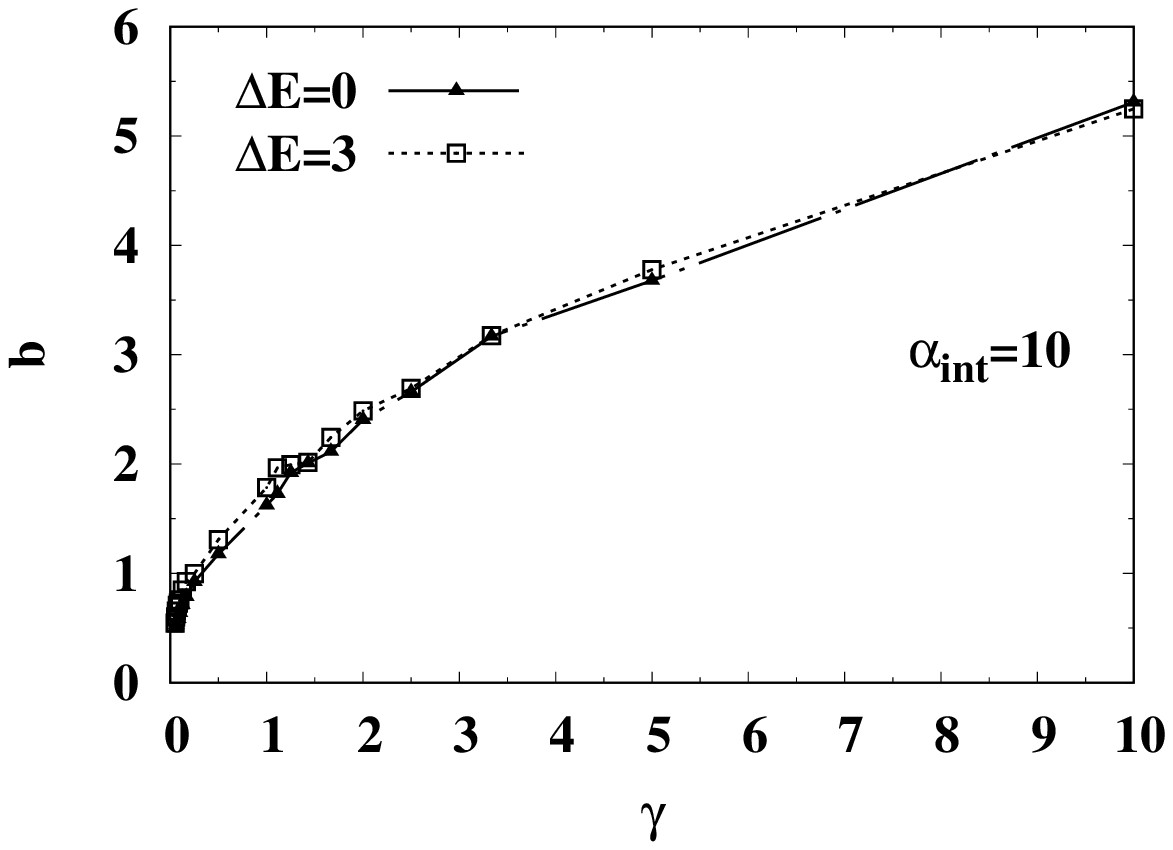}} \\
\end{minipage}
\caption{Dependence of $b$ on $\gamma$ at time $t = 2000$ for different values of the interaction parameter $\alpha_{int}$ and energy detuning $\Delta E$.  Parameter values: $\hbar=1, \gamma_R=0.1, \rho_0=1000,  V=\sqrt{0.1}/\rho_0 $.
} 
\label{Fig:NpF}
\end{figure}

Nevertheless, we can replace $b$ by a combined amplitude variable,
\begin{equation}
\beta = bm.
 \label{beta}
\end{equation}
Then it turns out that coupled equations
\begin{equation}
    \begin{cases}
    \frac{d\beta}{dt} = 
   m^{2}V^{2}\rho^{2}\cos{\theta}+\gamma\left\{\left(\frac{\beta}{m}\right)^{2}\cos{\theta}-\beta\right\}=0,
    \\
    \frac{dm}{dt} = 
  \gamma(\frac{\beta}{m}\cos{\theta}-m)=0,
    \\
    \frac{d\theta}{dt} =
    \frac{\alpha_{\text{int}}}{\hbar}\left(\frac{\beta}{m}\right)^{2}-V^{2}\rho^{2}\frac{m^2}{\beta}\sin{\theta}-\gamma\frac{\beta}{m^{2}}\sin{\theta}-\frac{\Delta E}{\hbar}=0,
     \\
     \frac{d\rho}{dt} =
     2\gamma(\rho_{0}-\rho)-2\beta V^{2}\rho^{2}\cos{\theta}=0
    \end{cases}
    \label{system}
\end{equation}
have non-trivial roots.
Since $b\to \infty$ with $t\to\infty$, time independence of $\beta$ means $m\rightarrow 0$. Then solution of (\ref{system}) is
\begin{equation}
    \beta_{\text{st}}=\gamma\frac{\hbar}{\alpha_{\text{int}}},\quad
        \theta_{\text{st}}=\frac{\pi}{2},\quad
        \rho_{\text{st}}=\rho_{0}.
        \label{focus}
\end{equation}
This fixed point is a stable focus that attracts all trajectories. It is illustrated in Fig.~\ref{Fig:beta_t} representing ensemble-averaged
time dependence of $\beta$. Fluctuations give rise to stochastic oscillations in the vicinity of $\beta=\beta_{\text{st}}$.
It is somewhat surprising that impact of fluctuations increases with decreasing $\gamma$ being decay rate of fluctuation correlations.
This circumstance has a simple explanation: increasing of $\gamma$ enhances growth of the oscillator population thereby reducing effect of fluctuations which 
basically have limited amplitude.
Dependence of $b$ on $\gamma$ at time instant $t=2000$ is presented in Fig.~\ref{Fig:NpF}.

\begin{figure}[!ht]
\centering
\includegraphics[width=.32\textwidth]{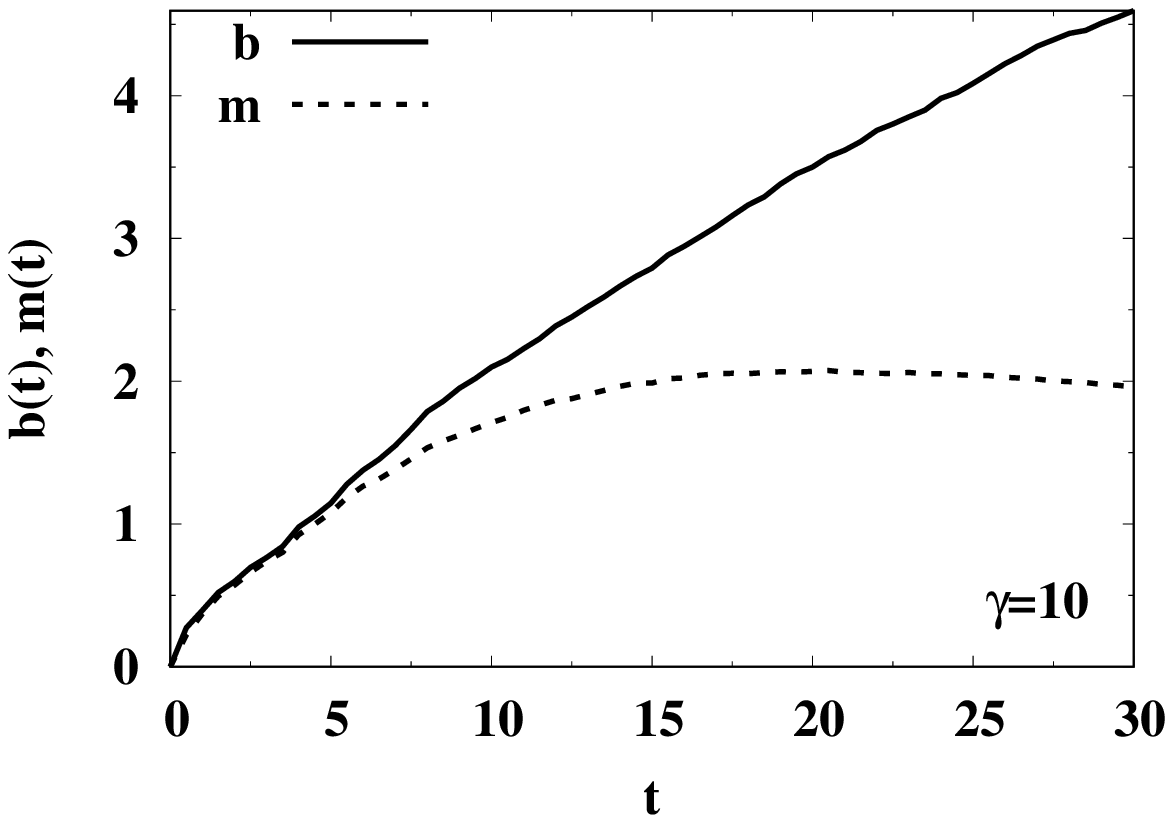}
\includegraphics[width=.32\textwidth]{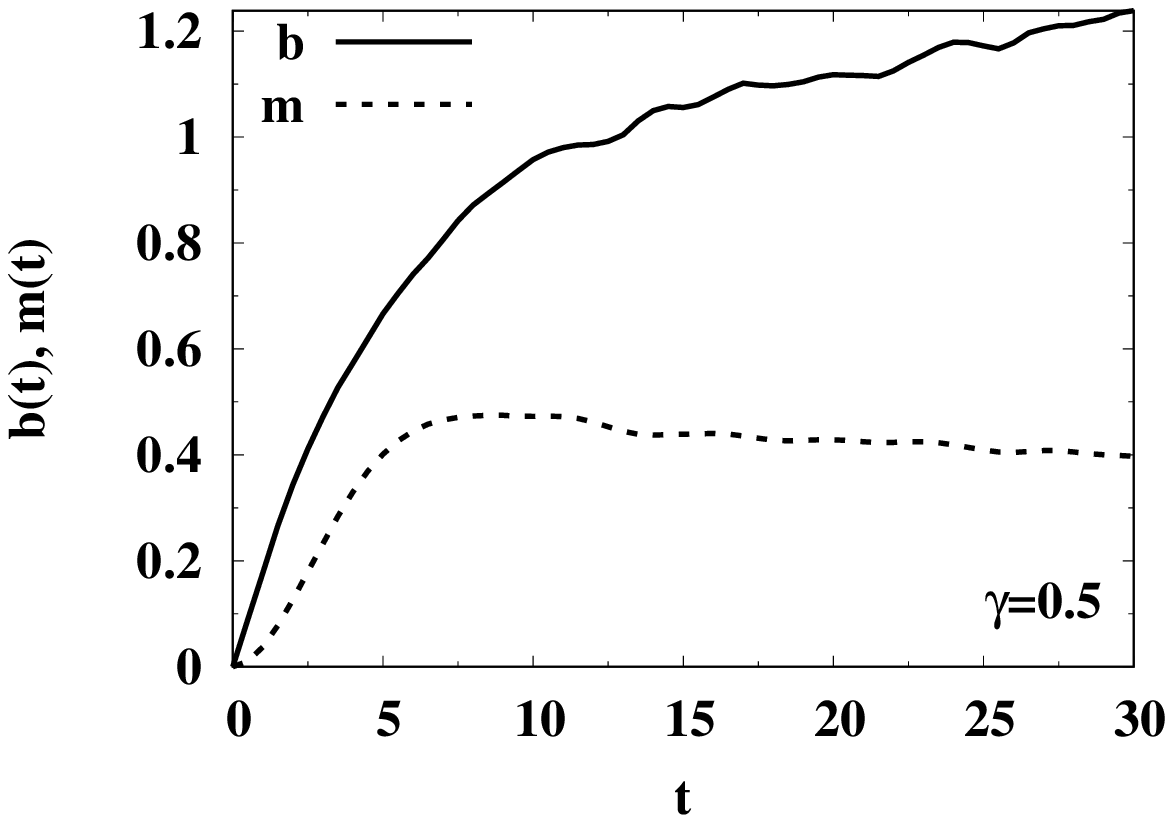}
\includegraphics[width=.32\textwidth]{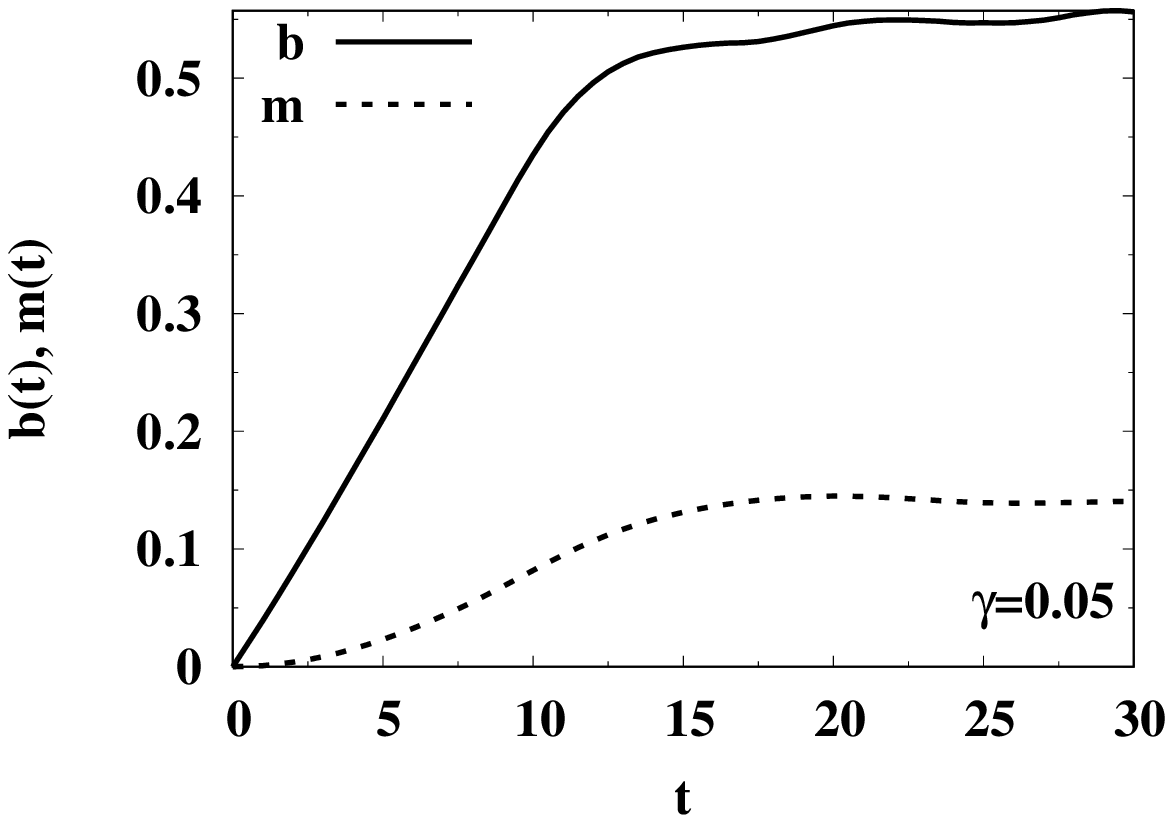}
\caption{Initial stage of the oscillator evolution: simultaneous growth of
oscillator amplitude $b$ and memory amplitude $m$. All the curves correspond to averaging  over 100 realizations of fluctuations.  Parameter values: $\hbar=1, \alpha_{\text{int}}=1, \gamma_R=0.1, \Delta E=0, \rho_0=1000, \Gamma=0.1, V=\sqrt{\Gamma}/\rho_0 $.
} 
\label{Fig:bm_t}
\end{figure}

Taking into account that the Markovian regime anticipates $M=a$, $\theta = 0$ and,
consequently, $\beta\to\infty$, one can conclude that the stationary point (\ref{focus})
is essentially non-Markovian. On the other hand, efficient pumping of the oscillator
implies nearly in-phase variations of $a$ and $M$ at least within some limited time interval.
This regime should be especially pronounced in the case of large values of $\gamma$ which
correspond to fast growth of oscillator amplitude (see Fig.~\ref{Fig:b_t}).
Data presented in Figure \ref{Fig:bm_t} confirms this hypothesis. 
In the case of relatively fast decoherence rate, $\gamma=10$,
we observe almost completely Markovian behavior until $t\simeq 5$.
Simultaneous growth of $b$ and $m$ is also observed for $\gamma=0.5$ and
$\gamma=0.05$, but the corresponding curves start diverging immediately at $t=0$.
It means that phase shift $\theta$ between oscillator complex amplitude $a$ and memory variable
$M$ is accumulated very rapidly.
As $\theta$ approaches $\pi/2$, the memory amplitude starts decreasing: oscillator state falls into some close vicinity of the asymptotic fixed point (\ref{focus}).
At this moment growth of the oscillator population is significantly decelerated.
Difference between the stage of initial pumping and the stage of attraction to the focus
is notably pronounced in the case of $\gamma=0.05$, i.~e. for low decoherence rate (see Fig.~\ref{Fig:bm_t}(c)).

\section{Coherence properties}
\label{sec:Coherence}

The model (\ref{sys0}) can be used for studying formation of Bose-Einstein condensate in course of stimulated relaxation of particles from the reservoir.
Emergence of condensate implies high degree of coherence in many-body dynamics.
Coherence of a state can be quantified by means of the first- and second-order correlation functions defined as \cite{Schwendimann,Whittaker}
\begin{equation}
    g^{(1)}(\tau)=\frac{\left<a^{*}(t)a(t+\tau)\right>}{\left<|a(t)|^{2}\right>},
\end{equation}
and
\begin{equation}
        g^{(2)}(\tau)=\frac{\left<a(t)a(t+\tau)a^{*}(t)a^{*}(t+\tau)\right>}{\left<|a(t)|^{2}\right>^{2}},
    \end{equation}
respectively.

\begin{figure}[!ht]
\centering
\includegraphics[width=.49\linewidth]{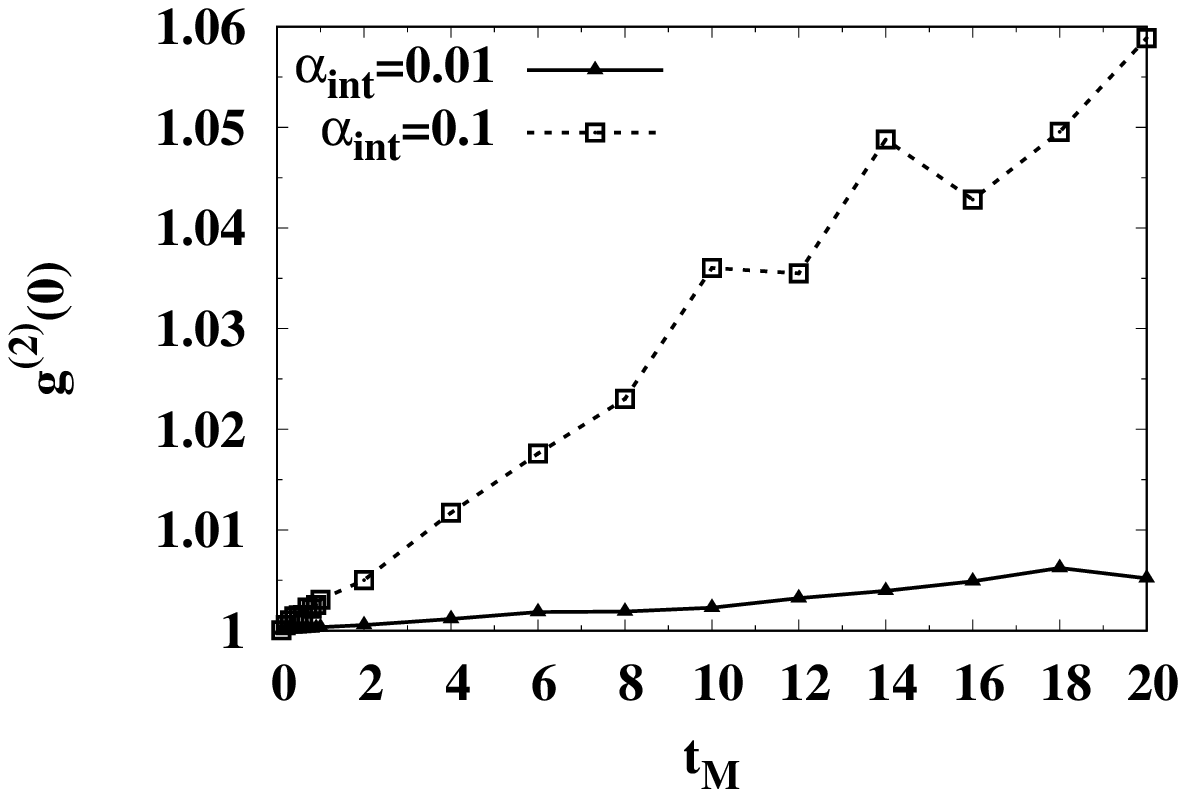}  
\includegraphics[width=.49\linewidth]{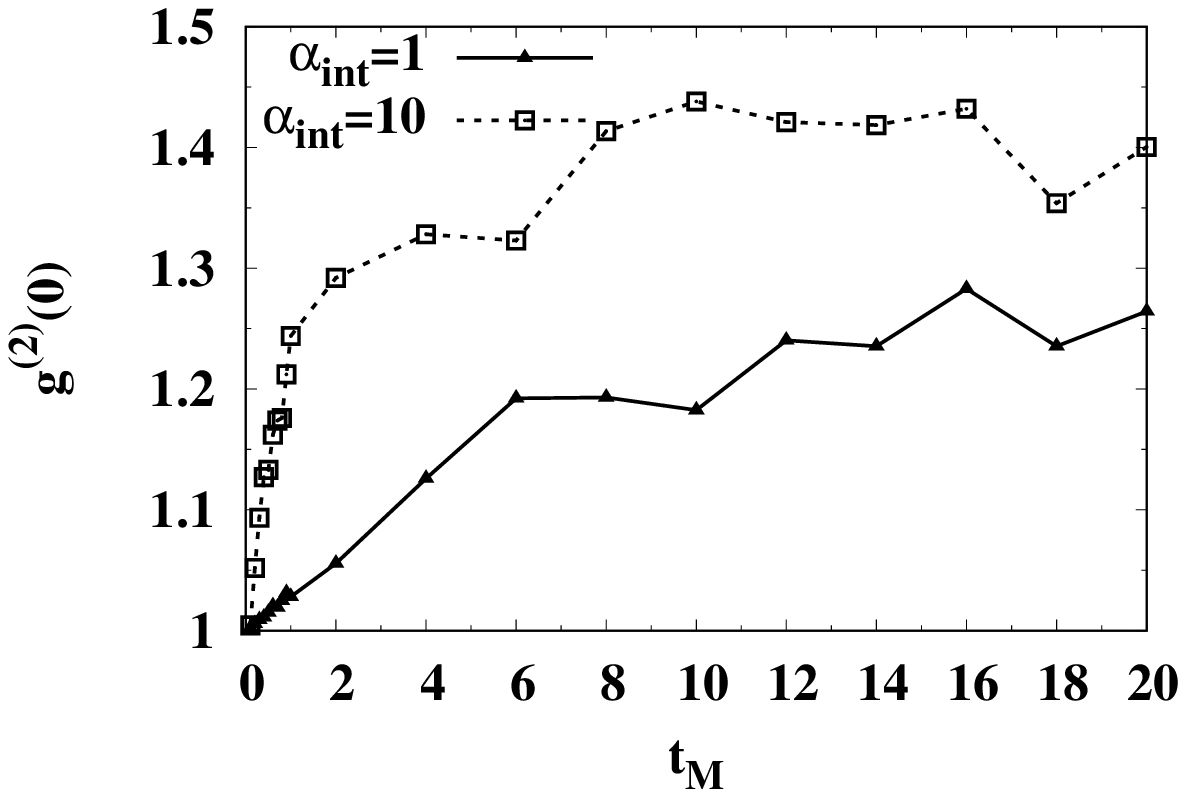} 
\caption{Dependence of $g^{(2)}(0)$ on $t_M$ for different values of the particle interaction parameter $\alpha_{\text{int}}$.  Parameter values: $\hbar=1, \gamma_R=0.1, \Delta E=0, \rho_0=1000, V=\sqrt{0.1}/\rho_0 $.
}
\label{Fig:g2}
\end{figure}

\begin{figure}[!ht]
\centering
\includegraphics[width=.32\textwidth]{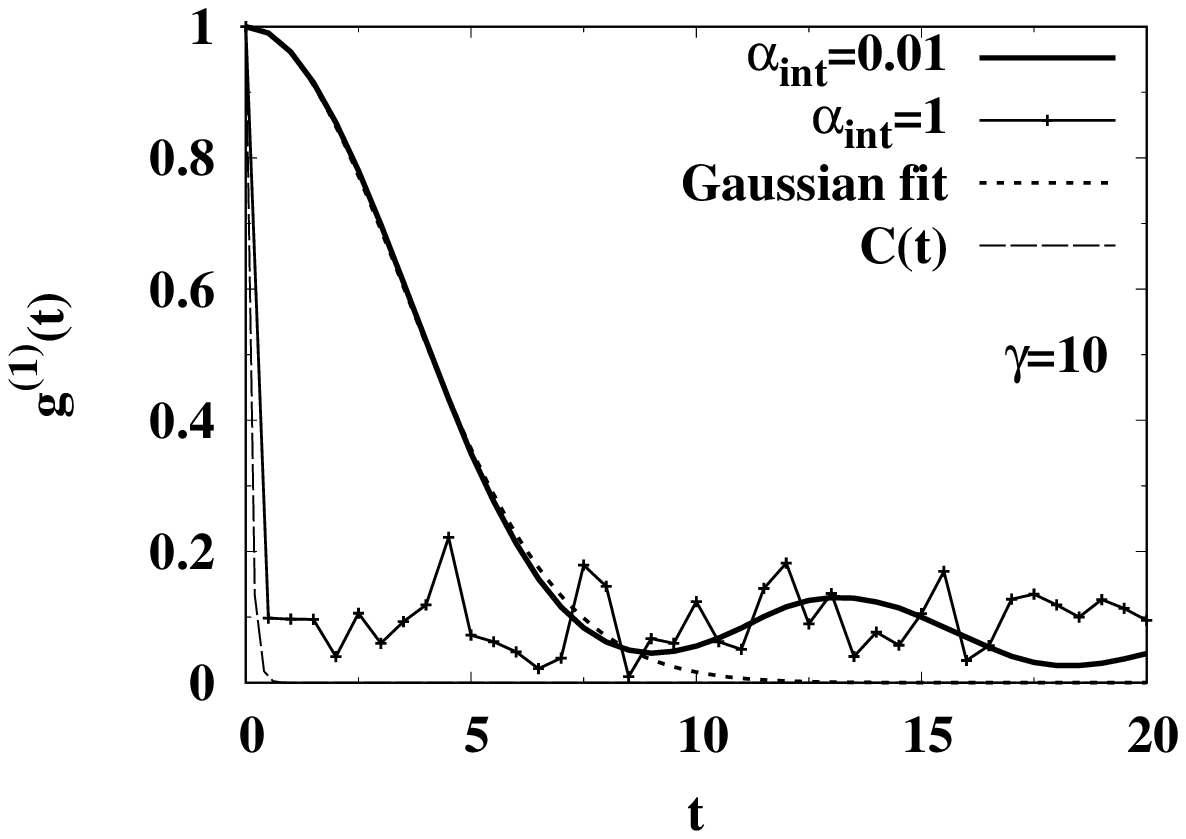}
\includegraphics[width=.32\textwidth]{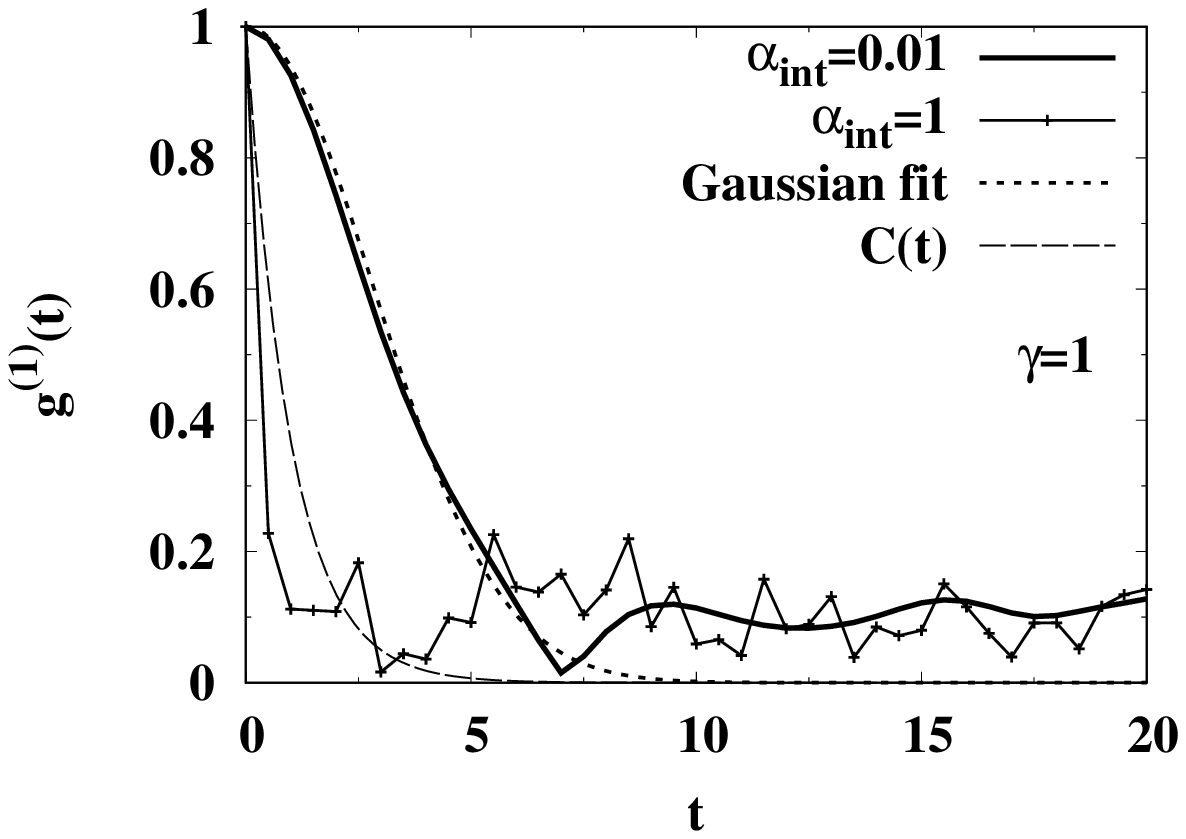}
\includegraphics[width=.32\textwidth]{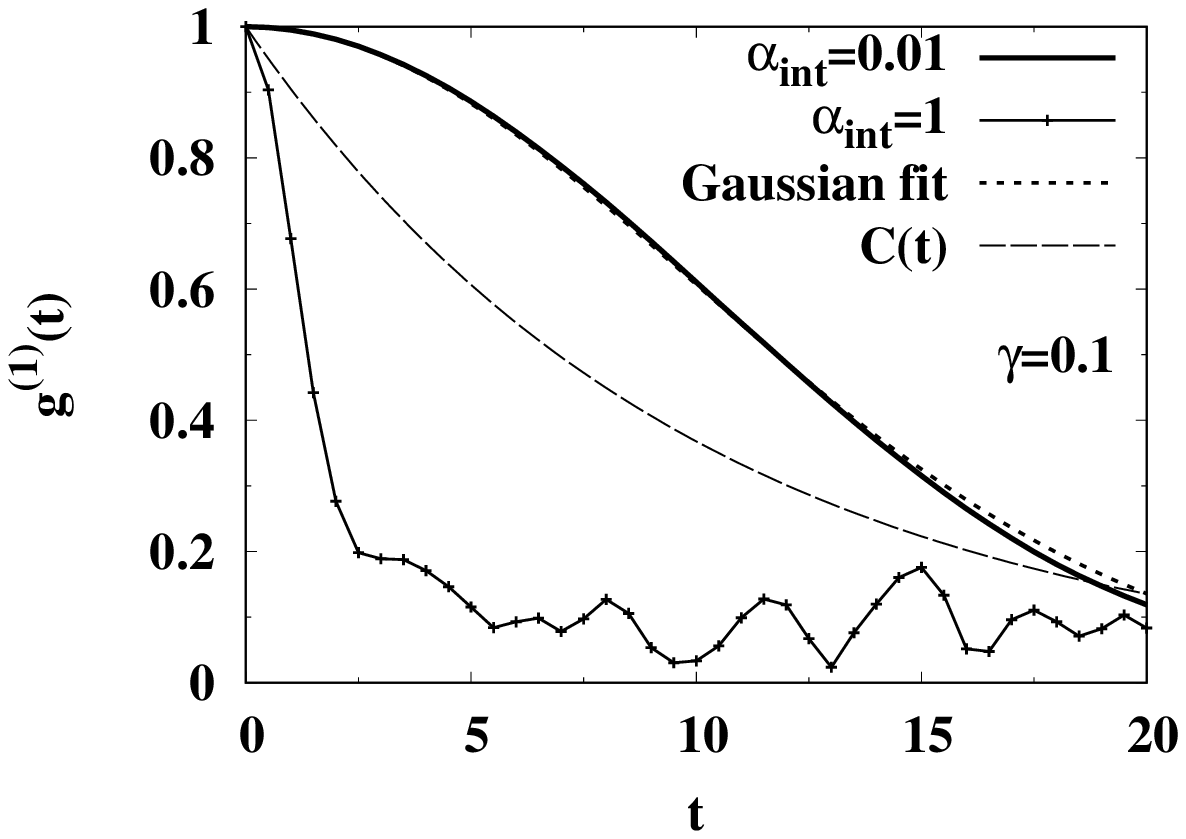}
\caption{Dependence of the first-order correlation function $g^{(1)}(t)$ on time for different values $\alpha_{\text{int}}$ and $\gamma$. $C(t)$  is the normalized autocorrelation function of the fluctuations.  Parameter values: $\hbar=1, \gamma_R=0.1, \Delta E=0, \rho_0=1000, V=\sqrt{0.1}/\rho_0 $.
}
\label{Fig:g1}
\end{figure}

The latter one describes population fluctuations.     
Its value for $\tau=0$ can be used as an indicator of condensate onset. 
The minimal value of $g^{(2)}$ is 1,
it corresponds to a many-body state being superposition of coherent single-particle states.
Deviation from 1 can be used as measure of decoherence. In the case of a thermal state we have $g^{(2)}(\tau=0)=2$.
Fig.~\ref{Fig:g2} represents dependence of $g^{(2)}(\tau=0)$ on memory time 
\begin{equation}
 t_{\text{M}} = \frac{1}{\gamma}
\end{equation}
for various values of the interaction strength $\alpha_{\text{int}}$. 
When calculating $g^{(1)}$ and $g^{(2)}$, we  considered only dynamics in the close vicinity of the focus point (\ref{focus}), omitting the transient early stage.
We see that impact of incoherence has tendency to increase with increasing $t_{\text{M}}$.
Comparing this plot with Fig.~\ref{Fig:b_t}, we can deduce that enhanced incoherence is associated with decreasing of state population.
Indeed, highly populated state is insensitive to fluctuations whose amplitude is limited and small.
Fluctuations can cause significant effect only onto a weakly-populated state.
Increasing of of interaction strength $\alpha_{\text{int}}$ also results in  diminishing of population and enhancing of decoherence.
So, we can conclude that state population is the main factor that determines value of $g^{(2)}(\tau=0)$.
It means that favourable conditions for formation
of Bose-Einstein condensate are expected for weakly interacting particles and broad reservoir spectrum,
the latter corresponds to large values of memory decay $\gamma$. 
On the other hand, analysis of data for $g^{(1)}(\tau)$ shows that condensate created with large values
of $\gamma$ undergoes faster decay of phase correlations, i.~e. it has lesser lifetime (see Fig.~\ref{Fig:g1}).
Anyway, decreasing of interaction strength gives rise to condensates with relatively long times of correlation decay,
and this decay is Gaussian, as it follows from the data presented in Fig.~\ref{Fig:g1}.

In contrast, the regime of relatively strong interaction, $\alpha_{\text{int}}=1$, corresponds to a depleted state with 
nearly exponential correlation decay indicating dominance of fluctuations.
Transition from exponential to Gaussian decay indicates the transformation of oscillator spectrum from Lorentzian to Gaussian.
According to the Kubo stochastic line shape theory \cite{Kubo},
it could be considered as a signature of fluctuations suppression.

According to Fig.~\ref{Fig:g1}, decay of $g^{(1)}(\tau)$ for $\alpha_{\text{int}}=1$ and $\gamma=0.1$
is much faster than decay of normalized noise autocorrelation function given by
\begin{equation}
    C(\tau)=e^{-(\gamma+i\frac{\Delta E}{\hbar})\tau}.
\end{equation}
In this case relatively strong particle interaction compensates low state population, keeping nearly the same impact of nonlinearity as for highly
populated states corresponding to $\alpha_{\text{int}}=0.01$.
We suggest that decoherence amplifies due to interplay of nonlinearity and noise with long correlation time ($t_{\text{M}}=10$).
Strong sensitivity to fluctuations in this case can be a signature of dynamical chaos.
In fact, the presence of a stable focus impedes exponential divergence of trajectories. It means there can be only a specific kind of chaos termed
``weak chaos'' \cite{Zaslavsky,Nicolis} can occur in this case.

\section{Summary}

In the present work we address the issue of non-Markovian manifestations
in dynamics of a single quantum state pumped from a reservoir.
It is assumed that this is a many-body state, taking into account interactions
between particles in the mean-field treatment.
We identify the Markovian behavior as equality of oscillator amplitude $a$ and the corresponding
memory variable $M$, when the corresponding equations of motion become local in time. A weaker definition of nearly Markovian dynamics could imply in-phase variations of $a$ and $M$.
In fact, we found the stationary oscillator state corresponding to
\begin{equation}
 bm = \text{const}.
\end{equation}
As long as Markovian dynamics anticipates $m = b$ and exponential growth of both these amplitudes, we can conclude that this state 
is intrinsically non-Markovian. It is a stable focus that
determines long-term dynamics for all finite values of memory decay rate $\gamma$.
Consequently there is no transition 
between non-Markovian and Markovian regimes in the parameter space. However, if the reservoir is sufficiently ``hot'' and characterized by broad energy
spectrum, then the oscillator can exhibit Markovian behavior within early stage of evolution, until dynamical memory is accumulated. 
Duration of this stage increases with increasing of reservoir temperature. If it exceeds all relevant physical time scales, 
then the time-local Markov approximation should be accurate.

Another point to be emphasized is the role of particle interaction in formation of 
Bose-Einstein condensate. Our results show that the interaction diminishes
state population, that results in stronger impact of fluctuations which destroy coherence.
On the other hand, the interaction plays the major role in thermalization and
stimulated scattering of particles from reservoir to condensate, so it should facilitate condensate emergence.
It suggests that there should be some intermediate range of interaction strength
that corresponds to optimal conditions for condensate formation.

\section*{Acknowledgments}
\pst
This work was carried out within the State Task No.~121021700341-2 for the POI FEB RAS.

\end{document}